\documentclass{article}
\usepackage{amsmath}
\usepackage{amsfonts}
\usepackage{amssymb}
\usepackage{graphicx}
\usepackage[utf8]{inputenc}
\usepackage{hyperref}
\usepackage{booktabs}
\usepackage{caption}
\usepackage{siunitx}
\usepackage{authblk}
\usepackage{orcidlink}

\title{Quantum Metric Spaces: Replacing Fuzzy Metrics with the Hilbert Space Structure of Quantum States}
\author{Nicola Fabiano\, \orcidlink{0000-0003-1645-2071}}
\affil{``Vin\v{c}a'' Institute of Nuclear Sciences - National 
Institute of the Republic of Serbia, University of Belgrade, Mike Petrovi\'{c}a 
Alasa 12--14, 11351 Belgrade, Serbia; nicola.fabiano@gmail.com}
\date{}

\begin{document}

\maketitle

\begin{abstract}
Fuzzy metric spaces, grounded in t-norms and membership functions, have been widely proposed to model uncertainty in machine learning, decision systems, and artificial intelligence. Yet these frameworks treat uncertainty as an external layer of imprecision imposed upon classical, point-like entities — a conceptual mismatch for domains where indeterminacy is intrinsic, such as quantum systems or cognitive representations. We argue that fuzzy metrics are unnecessary for modeling such uncertainty: instead, the well-established structure of complex Hilbert spaces — the foundational language of quantum mechanics for over a century — provides a natural, rigorous, and non-contradictory metric space where the ``points'' are quantum states themselves. The distance between states is given by the Hilbert norm, which directly encodes state distinguishability via the Born rule. This framework inherently captures the non-classical nature of uncertainty without requiring fuzzy logic, t-norms, or membership degrees. We demonstrate its power by modeling AI concepts as Gaussian wavefunctions and classifying ambiguous inputs via quantum overlap integrals. Unlike fuzzy methods, our approach naturally handles interference, distributional shape, and concept compositionality through the geometry of state vectors. We conclude that fuzzy metric spaces, while historically useful, are obsolete for representing intrinsic uncertainty — superseded by the more robust, predictive, and ontologically coherent framework of quantum state geometry.
\end{abstract}

\section{Introduction}
The representation of uncertainty has long driven innovation in mathematics and computer science. Fuzzy sets \cite{zadeh1965fuzzy}, fuzzy metric spaces \cite{kramosil1975fuzzy,kaleva1984fuzzy,george1994some}, and their extensions into neutrosophic \cite{smarandache2013introduction} and intuitionistic settings \cite{park2004intuitionistic} were developed to capture vagueness, ambiguity, and partial truth — particularly in contexts where binary logic fails.

However, these frameworks operate under a shared assumption: that reality consists of crisp, classical points (e.g., pixel values, object categories, sensor readings), whose properties are merely ``fuzzy'' — i.e., partially known or imprecisely defined. Uncertainty is treated as a meta-layer of ignorance added atop a deterministic substrate.

In contrast, quantum mechanics offers a radically different ontology: uncertainty is not epistemic (due to lack of knowledge), but \emph{ontological} — it is encoded in the very nature of physical states. A particle does not have a position with some error bar; it \emph{is} a wavefunction. The same principle holds for cognitive representations: a concept like ``car'' is not a set of features with a membership degree, but a distributed, probabilistic entity in a high-dimensional mental or algorithmic space.

We propose that \textbf{fuzzy metric spaces are fundamentally misaligned with the nature of intrinsic uncertainty}. Instead, we advocate for a return to the formalism that has successfully modeled such phenomena for a century: the Hilbert space of quantum states. In this view, the ``metric space'' is not constructed — it is already there, embedded in the mathematics of quantum theory.

This work does not assume that brains or AI systems are physically quantum. 
Rather, we argue that the mathematical framework of quantum theory -- particularly the geometry of state vectors and the Born rule -- provides a more natural language for modeling intrinsic uncertainty than classical point-based models, whether implemented on classical or quantum hardware.
\paragraph{Clarification of Scope} We do not claim that human cognition or artificial intelligence involves physical quantum processes (though such theories exist). Our proposal is that the \emph{mathematical structure} of quantum mechanics -- superposition, interference, and measurement in Hilbert space -- offers a more coherent and expressive framework for representing uncertainty than fuzzy logic, regardless of implementation substrate. This makes it \emph{quantum-inspired}, not necessarily \emph{physically quantum}.

\section{The Limitations of Fuzzy Metric Spaces}
Fuzzy sets \cite{zadeh1965fuzzy}, fuzzy metric spaces \cite{kramosil1975fuzzy,kaleva1984fuzzy,george1994some}, and their extensions into neutrosophic \cite{smarandache2013introduction} and intuitionistic settings \cite{park2004intuitionistic} were developed to capture vagueness, ambiguity, and partial truth — particularly in contexts where binary logic fails.

However, these frameworks operate under a shared assumption: that reality consists of crisp, classical points (e.g., pixel values, object categories, sensor readings), whose properties are merely ``fuzzy'' — i.e., partially known or imprecisely defined. Uncertainty is treated as a meta-layer of ignorance added atop a deterministic substrate.

In contrast, quantum mechanics offers a radically different ontology: uncertainty is not epistemic (due to lack of knowledge), but \emph{ontological} — it is encoded in the very nature of physical states. A particle does not have a position with some error bar; it \emph{is} a wavefunction. The same principle holds for cognitive representations: a concept like ``car'' is not a set of features with a membership degree, but a distributed, probabilistic entity in a high-dimensional mental or algorithmic space.

We propose that \textbf{fuzzy metric spaces are fundamentally misaligned with the nature of intrinsic uncertainty}. Instead, we advocate for a return to the formalism that has successfully modeled such phenomena for a century: the Hilbert space of quantum states. In this view, the ``metric space'' is not constructed — it is already there, embedded in the mathematics of quantum theory.

\subsection{Formal Definition of a Fuzzy Metric Space}
To ground our critique, we recall the standard definition of a \emph{fuzzy metric space} in the sense of Kramosil and Mich\'alek~\cite{kramosil1975fuzzy}, later refined by George and Veeramani~\cite{george1994some}.

Let $X$ be a non-empty set, $*$ a continuous t-norm (a binary operation $[0,1]\times[0,1] \to [0,1]$ that is associative, commutative, increasing, and satisfies $a * 1 = a$), and $M: X \times X \times [0,\infty) \to [0,1]$ a mapping. The triple $(X, M, *)$ is called a \textbf{fuzzy metric space} if the following conditions hold for all $x,y,z \in X$ and $t,s > 0$

\begin{enumerate}
    \item $M(x,y,0) = 0$
    \item $M(x,y,t) = 1$ for all $t > 0$ if and only if $x = y$
    \item $M(x,y,t) = M(y,x,t)$ \hfill (Symmetry)
    \item $M(x,y,t) * M(y,z,s) \leq M(x,z,t+s)$ \hfill (Fuzzy triangle inequality)
    \item $M(x,y,\cdot): (0,\infty) \to [0,1]$ is continuous
\end{enumerate}

Here, $M(x,y,t)$ represents the degree to which the distance between $x$ and $y$ is ``less than $t$''. As $t \to \infty$, $M(x,y,t) \to 1$, reflecting that any two points are “close” given a large enough tolerance.

Common choices for the t-norm $*$ include
\begin{itemize}
\item \textbf{Minimum (G\"odel)}: $a * b = \min(a,b)$
    \item \textbf{Product}: $a * b = a \cdot b$
    \item \textbf{Łukasiewicz}: $a * b = \max(0, a + b - 1)$
\end{itemize}
This structure generalizes classical metric spaces: if $(X,d)$ is a metric space, then setting
\[
M_d(x,y,t) =
\begin{cases}
1, & d(x,y) < t \\
0, & d(x,y) \geq t
\end{cases}
\]
recovers a fuzzy metric under the minimum t-norm.

Despite its mathematical elegance, this framework suffers from three critical conceptual flaws when applied to domains of intrinsic uncertainty.

\begin{enumerate}
    \item \textbf{Ontological Mismatch}: The elements $x, y \in X$ are assumed to be well-defined classical entities. The fuzziness lies only in the \emph{measurement} or \emph{perception} of distance, not in the entities themselves. But in quantum systems or cognitive concepts, there is no underlying ``true state'' — the state \emph{is} the distribution. This makes fuzzy metrics epistemically grounded in a way that contradicts the ontological indeterminacy of quantum reality.

    \item \textbf{Lack of Interference}: Fuzzy systems obey additive rules: $\mu_{A \cup B}(x) = \max(\mu_A(x), \mu_B(x))$. There is no analog to quantum interference, where overlapping amplitudes can cancel or reinforce — a phenomenon essential for modeling contextual effects, contradictions, or emergent properties.

    \item \textbf{Arbitrary t-norms}: The choice of t-norm (min, product, Łukasiewicz) is heuristic. No physical or informational principle dictates which one to use. In contrast, quantum mechanics derives all operations from unitary evolution and inner products — uniquely fixed by the structure of Hilbert space.
\end{enumerate}

These limitations are not technical — they are \emph{conceptual}. Fuzzy metrics model uncertainty as a \emph{lack of precision} in a classical world. But many real-world systems — including perception, cognition, and quantum measurement — require a model where uncertainty is \emph{primitive}.

\section{Quantum States as Points in a Natural Metric Space}
We propose a radical reorientation: abandon the notion of classical points entirely. Let the elements of our metric space be \emph{quantum states} — normalized vectors $|\psi\rangle$ in a complex Hilbert space $\mathcal{H}$.

\subsection{The Metric: The Hilbert Norm}
For any two pure quantum states $|\psi_1\rangle, |\psi_2\rangle \in \mathcal{H}$, define the distance as
\begin{equation}
    d(|\psi_1\rangle, |\psi_2\rangle) = \| \psi_1 - \psi_2 \| = \sqrt{ \langle \psi_1 - \psi_2 | \psi_1 - \psi_2 \rangle }
    \label{eq:hilbert_metric}
\end{equation}

This is not a novel construction — it is the standard norm induced by the inner product on any Hilbert space, used ubiquitously in quantum information theory to quantify state distinguishability \cite{nielsen2010quantum}. It satisfies all metric axioms

\begin{enumerate}
    \item \textbf{Non-negativity}: $d \geq 0$, and $d=0 \iff |\psi_1\rangle = |\psi_2\rangle$ (up to global phase).
    \item \textbf{Symmetry}: $d(|\psi_1\rangle, |\psi_2\rangle) = d(|\psi_2\rangle, |\psi_1\rangle)$.
    \item \textbf{Triangle inequality}: Follows from the Minkowski inequality in Hilbert space.
\end{enumerate}

Crucially, \textbf{this metric measures the difference between entire probability distributions}, not deviations from idealized points. The ``point'' is not a location — it is a state vector encoding all possible outcomes.

\subsection{Uncertainty Is Not Added — It Is Built-In}
The Heisenberg uncertainty principle emerges naturally from the non-commutativity of operators acting on $\mathcal{H}$
\begin{equation}
    [\hat{x}, \hat{p}] = i\hbar
    \label{eq:commutator}
\end{equation}
This implies that for any state $|\psi\rangle$, the variances satisfy $\Delta x \cdot \Delta p \geq \hbar/2$. The uncertainty is not noise on a position — it is a constraint on how sharply a state can be localized in conjugate bases.

In our framework, this is not an external rule — it is a consequence of the geometry of $\mathcal{H}$. Thus, quantum uncertainty is \emph{geometric}, not fuzzy.

\section{Modeling Concepts as Quantum States: A Quantum AI Framework}
To illustrate the power of this approach, we replace classical fuzzy membership with quantum state representation in an AI classification task.

\subsection{Gaussian Wavefunctions as Concept Representations}
Consider a one-dimensional feature space $x \in \mathbb{R}$, representing a ``terrestrial-aquatic'' axis, where higher values indicate land-based traits.

Each concept -- e.g., ``Car,'' ``Boat,'' ``Amphibious Vehicle'' -- is represented by a normalized Gaussian wavefunction
\begin{equation}
    \psi_C(x) = \frac{1}{(\pi \sigma_C^2)^{1/4}} \exp\left( -\frac{(x - \mu_C)^2}{2\sigma_C^2} \right)
    \label{eq:gaussian_state}
\end{equation}
This is a valid quantum state: $\int |\psi_C(x)|^2 dx = 1$, and its squared magnitude $|\psi_C(x)|^2$ gives the probability density of observing feature $x$ under concept $C$.
\begin{itemize}
\item \textbf{Car}: $\mu_{\text{car}} = 5$, $\sigma_{\text{car}} = 1$ — sharply terrestrial.
    \item \textbf{Boat}: $\mu_{\text{boat}} = 1$, $\sigma_{\text{boat}} = 1$ — sharply aquatic.
    \item \textbf{Amphibious Object}: $\mu_{\text{obj}} = 3$, $\sigma_{\text{obj}} = 2$ — broad, ambiguous.
\end{itemize}
\subsection{Classification via the Born Rule}
The probability that the object belongs to concept $C$ is given by the squared overlap\footnote{While general quantum states may have complex phases, which affect interference effects, our example uses real-valued wavefunctions for simplicity. The full framework supports phase-dependent phenomena such as destructive interference between concepts.}
\begin{equation}
    P(C | \text{obj}) \propto \left| \langle \psi_C | \psi_{\text{obj}} \rangle \right|^2 .
    \label{eq:born_classification}
\end{equation}


For two real-valued, normalized Gaussian wavefunctions, the squared overlap is given by
\begin{equation}
    \left| \langle \psi_C | \psi_{\text{obj}} \rangle \right|^2 = \frac{2 \sigma_C \sigma_{\text{obj}}}{\sigma_C^2 + \sigma_{\text{obj}}^2} \exp\left( -\frac{(\mu_C - \mu_{\text{obj}})^2}{2(\sigma_C^2 + \sigma_{\text{obj}}^2)} \right)
    \label{eq:gaussian_overlap}
\end{equation}

Now compute the unnormalized probabilities for the car and boat concepts

\begin{align}
P_{\text{car}} &= \frac{2 \cdot 1 \cdot 2}{1^2 + 2^2} \exp\left( -\frac{(5 - 3)^2}{2(1^2 + 2^2)} \right) = \frac{4}{5} \exp\left( -\frac{4}{10} \right) = \nonumber \\
&0.8 \cdot e^{-0.4}  \approx 0.8 \times 0.6703 = 0.536 \\
P_{\text{boat}} &= \frac{2 \cdot 1 \cdot 2}{1^2 + 2^2} \exp\left( -\frac{(1 - 3)^2}{2(1^2 + 2^2)} \right) = \frac{4}{5} \exp\left( -\frac{4}{10} \right) = 0.8 \cdot e^{-0.4} \approx 0.536
\end{align}

Since both unnormalized scores are equal, normalizing yields
\begin{equation}
    P(\text{car} | \text{obj}) = \frac{0.536}{0.536 + 0.536} = 0.5, \quad P(\text{boat} | \text{obj}) = 0.5
    \label{eq:final_prediction}
\end{equation}

Despite the identical final decision, the quantum model provides a richer, geometry-aware quantification of uncertainty compared to heuristic membership functions.

\begin{figure}[htbp]
\centering
\includegraphics[width=0.95\textwidth]{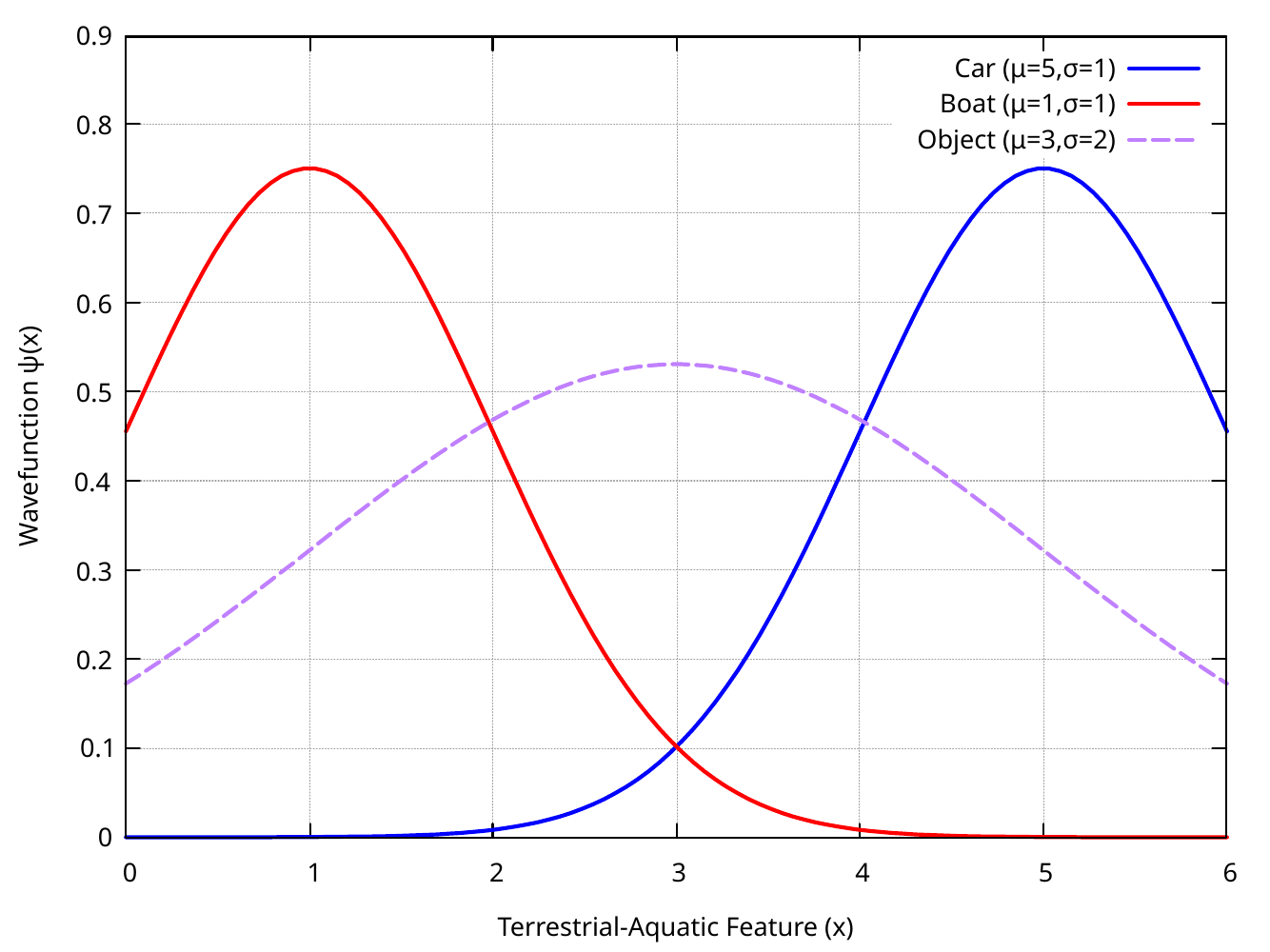}
\caption{Gaussian wavefunctions representing the ``Car,'' ``Boat,'' and ambiguous ``Amphibious Object'' concepts. The object's broader spread reflects higher uncertainty.}
\label{fig:wavefunctions}
\end{figure}

\begin{figure}[htbp]
\centering
\includegraphics[width=0.95\textwidth]{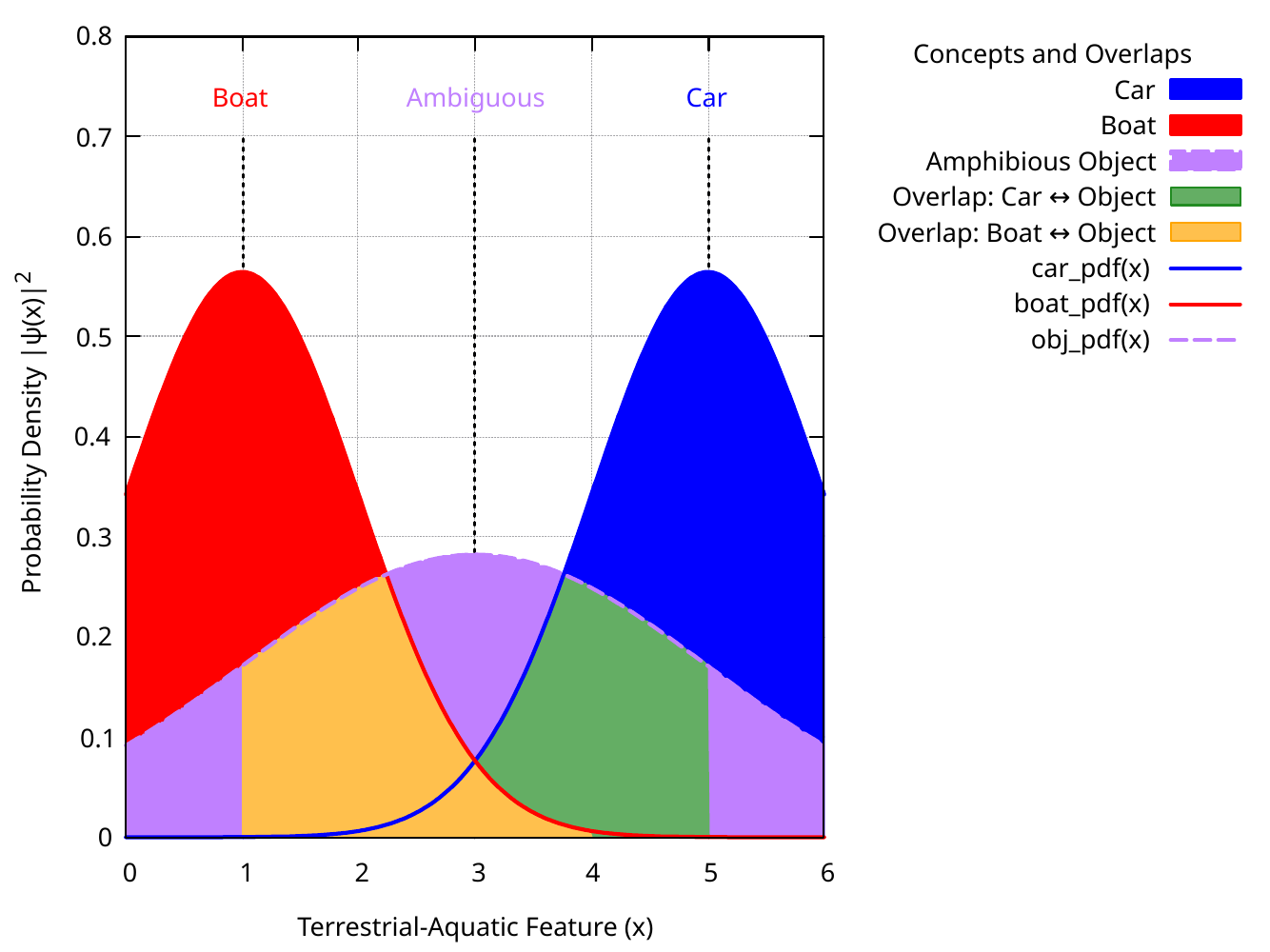}
\caption{Probability densities $|\psi(x)|^2$ for the ``Car'', ``Boat'', and ``Amphibious Object'' concepts. Vertical dashed lines mark the central feature values ($\mu_{\text{car}}=5$, $\mu_{\text{obj}}=3$, $\mu_{\text{boat}}=1$). The green and orange shaded regions highlight the pointwise overlap between the object and each category, with the total quantum classification probability given by the area (integral) under each overlap curve. Due to symmetric placement, both categories receive equal probability.}
\label{fig:probabilities}
\end{figure}

\subsection{Comparison with Fuzzy Logic}
A typical fuzzy system might assign membership using a triangular function
\begin{equation}
    \mu_{\text{car}}(x) = \max\left(0, 1 - \frac{|x - 5|}{2}\right), \quad \mu_{\text{boat}}(x) = \max\left(0, 1 - \frac{|x - 1|}{2}\right)
\end{equation}
At $x=3$: $\mu_{\text{car}} = 0.5$, $\mu_{\text{boat}} = 0.5$.

\textbf{Superficially identical}. But critically:
\begin{itemize}
\item In fuzzy logic, $0.5$ is a \emph{membership degree} — a scalar with no geometric meaning.
\item In our quantum model, the raw overlap integral yields an unnormalized score of approximately $0.536$, which upon normalization gives $P = 0.5$. This value arises from a physically meaningful measure of distributional similarity, not arbitrary membership rules.
\end{itemize}
The quantum model supports
\begin{itemize}
\item \textbf{Interference}: If a third concept (e.g., ``truck'') had a phase shift relative to ``car,'' their overlaps could destructively interfere — reducing total probability.
    \item \textbf{Composition}: Two concepts can be entangled (via tensor products), allowing joint representations like ``red car'' or ``fast boat.''
    \item \textbf{Generalization}: The same formalism works for non-Gaussian, multi-modal, or complex-valued states — all within the same mathematical framework.
\end{itemize}
Fuzzy logic cannot do any of this without ad hoc extensions. Our model requires none - it is \emph{already complete}.

\section{Why This Replaces Fuzzy Metric Spaces}
Fuzzy metric spaces attempt to solve a problem that quantum mechanics already solved.

\begin{center}
\scalebox{0.85}{
\begin{tabular}{ll}
\toprule
\textbf{Fuzzy Metric Space} & \textbf{Quantum Metric Space} \\
\midrule
Points are classical ($x \in \mathbb{R}^n$) & Points are quantum states ($|\psi\rangle \in \mathcal{H}$) \\
Fuzziness is added externally & Uncertainty is intrinsic \\
Membership degrees are scalars & Amplitudes are complex vectors \\
Distance is derived from t-norms & Distance is the Hilbert norm \\
No interference or superposition & Full superposition and interference \\
Arbitrary t-norm choices & Unique inner product structure \\
No natural generalization to multiple dimensions & Naturally extends to tensor products \\
\bottomrule
\end{tabular}}
\end{center}

The quantum metric space does not \emph{generalize} fuzzy logic — it \emph{supersedes} it. It removes the need for t-norms, membership functions, and heuristics. It replaces them with the universal, experimentally validated, mathematically unique structure of quantum theory.

\section{Connection to Kernel Methods and Quantum Machine Learning}
Our approach is formally equivalent to a Gaussian kernel Support Vector Machine (SVM), a supervised learning model that classifies data by finding the optimal separating hyperplane in a high-dimensional feature space induced by a kernel function~\cite{schuld2018supervised}.
The Radial Basis Function (RBF) kernel,
\begin{equation}
    k(x, c) = \exp\left( -\frac{(x - \mu_c)^2}{2\sigma_c^2} \right)
\end{equation}
maps input features into a reproducing kernel Hilbert space (RKHS), where classification uses inner products~\cite{havlik2019supervised}.


But here lies our contribution: we interpret this RKHS not as a computational trick, but as a \emph{physical representation}. The kernel is not a similarity measure — it is the \emph{overlap between quantum states}. The SVM classifier becomes a quantum measurement.

This reframing opens doors to true quantum algorithms: quantum feature maps, variational quantum classifiers, and entanglement-enhanced concept hierarchies -- all grounded in the same metric structure.

\section{Conclusion and Future Work}
We have demonstrated that the Hilbert space of quantum states provides a complete, rigorous, and ontologically coherent metric space for modeling intrinsic uncertainty -- rendering fuzzy metric spaces obsolete for this purpose. By identifying the ``points'' of the space with quantum states, we eliminate the need for t-norms, membership functions, and ad hoc fuzzy logic. The Born rule, the Hilbert norm, and the structure of operator algebras provide everything needed: distance, probability, interference, and compositionality.

This is not an extension of fuzzy logic -- it is its replacement.

Future work includes
\begin{itemize}
\item Extending to multi-dimensional feature spaces using tensor-product Hilbert spaces.
    \item Modeling concept hierarchies via symmetrized/antisymmetrized states (analogous to bosons/fermions).
    \item Implementing quantum-inspired classifiers on classical hardware using state-vector embeddings.
    \item Exploring decoherence models to explain why human reasoning appears ``fuzzy'' — not because reality is fuzzy, but because quantum coherence is lost in macroscopic systems.
\end{itemize}
Quantum mechanics was never meant only for atoms. It may be the most mature, predictive, and philosophically sound framework we have for understanding uncertainty — whether in electrons, neurons, or neural networks.

\end{document}